\documentclass{iau} 
\usepackage{graphicx}

\title[Dynamical models of stellar systems] 
      {Dynamical models of spheroidal multi-component stellar systems}

\author[C. Caravita, L. Ciotti and S. Pellegrini] 
	     {Caterina Caravita $^{1,2}$, Luca Ciotti $^{1}$ and Silvia Pellegrini $^{1,2}$}

\affiliation{$^{1}$ Department of Physics and Astronomy, University of Bologna, \\
            via P. Gobetti 93/2, 40129 Bologna, Italy \\
            email: {\tt caterina.caravita2@unibo.it} \\[\affilskip]
            $^{2}$ INAF-OAS of Bologna, \\
            via P. Gobetti 93/3, 40129 Bologna, Italy}

\pubyear{2019}
\volume{351}  
\jname{Star Clusters: From the Milky Way to the Early Universe}
\editors{A. Bragaglia, M.B. Davies, A. Sills \& E. Vesperini, eds.}
\begin{document}

\maketitle

\begin{abstract}
We present a significantly improved version of our numerical code JASMINE, that can now solve the Jeans equations for axisymmetric models of stellar systems, composed of an arbitrary number of stellar populations, a Dark Matter halo, and a central Black Hole. The stellar components can have different structural (density profile, flattening, mass, scale length), dynamical (rotational support, velocity dispersion anisotropy), and population (age, metallicity, Initial Mass Function, mass-to-light ratio) properties. These models, when combined with observations, will allow to investigate important issues, such as quantifying the systematic effects of IMF variations, of mass-to-light ratio gradients, and of different stellar kinematic components (e.g. counter rotating disks, kinematically decoupled cores) on luminosity-weighted properties. The developed analytical and numerical framework aims at modeling Early-Type Galaxies, but it can also be applied to dwarf Spheroidal galaxies and Globular Clusters.

\keywords{methods: analytical, numerical; galaxies: elliptical and lenticular, dwarf, star clusters, stellar content, kinematics and dynamics} 
\end{abstract}

\firstsection 
\section{Jeans models}

In our study, we employ the JASMINE code (\cite[Posacki et al., 2013]{pos13}) to construct axisymmetric stellar systems, composed of stars, a Dark Matter halo (DM) and a central Black Hole (BH), through the solution of the Jeans Equations (JEs) in cylindrical coordinates. The total stellar distribution is modeled as the superposition of $i=1,...\,,N$ different components: from a structural point of view, each stellar component is defined by an assigned axisymmetric density distribution $\rho_{*i}$, with total mass $M_{*i}$ and possible additional physical scales (such as characteristic length $r_{*i}$, truncation radius, etc.) and parameters (such as flattening $q_i$, in case of oblate ellipsoidal component). A DM halo of density distribution $\rho_\mathrm{h}$ and total mass $M_\mathrm{h}$, and a central BH of mass $M_\mathrm{BH}$, can be included. The total gravitational potential due to the stars and the DM halo is numerically computed through the solution of the Poisson equation.

Each stellar density component of the resulting models, with symmetry axis aligned with the $z$-axis, is assumed to be described by a two-integral distribution function $f_i(E,J_\mathrm{z})$. As well known, under these assumptions, $\overline{v_\mathrm{R}}_i=\overline{v_\mathrm{z}}_i=0$, and the only non-zero rotational velocity can be in the azimuthal direction, $\overline{v_\varphi}_i$. In addition, the velocity dispersion tensor is diagonal and aligned with the coordinate system, and it can have only azimuthal anisotropy (while $\sigma_{\mathrm{R}i}=\sigma_{\mathrm{z}i}\equiv \sigma_i$). It follows that the JEs for the stellar distribution $\rho_{*i}$, embedded in a total gravitational potential $\Phi_\mathrm{tot}=\sum_i\phi_{*i}+\phi_\mathrm{h}+\phi_\mathrm{BH}$ (sum of the contributions of stars, DM and BH), are given by:
\begin{equation} \label{eq:vertJEs}
 \displaystyle{\frac{\partial \big(\rho_{*i}\sigma_i^2\big)}{\partial z} = -\rho_{*i} \frac{\partial \Phi_\mathrm{tot}}{\partial z}} \,,
\end{equation}
\begin{equation} \label{eq:radJEs}
 \displaystyle{\frac{\partial \big(\rho_{*i}\sigma_i^2\big)}{\partial R} = \frac{\rho_{*i} \Big(\overline{v_\varphi^2}_i-\sigma_i^2\Big)}{R} -\rho_{*i} \frac{\partial \Phi_\mathrm{tot}}{\partial R}} \,,
\end{equation}
(see e.g. \cite[Binney \& Tremaine, 2008]{b&t08}).

In order to split the azimuthal kinetic energy ($\overline{v_\varphi^2}_i$) between streaming and random motions, we adopt the $k$-decomposition (\cite[Satoh, 1980]{satoh80}), with the introduction of a constant parameter $k_i$, in principle different for each stellar component, so that
\begin{equation} \label{eq:satoh}
 \overline{v_\varphi}_i^2 = k_i^2 \Big(\overline{v_\varphi^2}_i-\sigma_i^2\Big) \,, \qquad
 \sigma_{\varphi i}^2 \equiv \overline{v_\varphi^2}_i - \overline{v_\varphi}_i^2  = \sigma_i^2 + \big(1-k_i^2\big) \Big(\overline{v_\varphi^2}_i-\sigma_i^2\Big) \,.
\end{equation}
Usually $0\leq k_i \leq 1$, but more general choices are possible (see e.g. \cite[Ciotti \& Pellegrini, 1996]{ciot-pel96}). If $k_i=1$, the component into consideration is an isotropic rotator ($\sigma_{\varphi i}=\sigma_i$) and it is fully rotationally supported; if $k_i=0$, there is no net rotation ($\overline{v_\varphi}_i=0$), so the component is supported only by the azimuthal anisotropy of the velocity dispersion tensor. In order to introduce anisotropy between the radial and vertical velocity dispersions (implicitly relaxing the assumption of a two-integral distribution function), i.e. to have $\sigma_{\mathrm{R}i}\neq\sigma_{\mathrm{z}i}$, the JASMINE code can be easily modified to include, for example, the Cappellari (\cite[2008]{capp08}) anisotropy.

Once the kinematic fields are derived, we construct their mass and luminosity-weighted projections along a given line-of-sight direction, and we obtain the corresponding 2D maps. In turn, these maps can be compared with the analogous maps of real stellar systems, as obtained for instance from Integral Field Spectroscopy.

The linearity of the JEs, with respect to $\rho_{*i}$ and $\Phi_\mathrm{tot}$, allows for a significant amount of \textit{scaling} of the solutions. Indeed, since the general solutions are built by adding the solutions pertaining to each stellar component $\rho_{*i}$, in the gravitational field produced by $\Phi_\mathrm{tot}$, it is possible to considerably reduce the amount of numerical work required. In practice, for a given set of mass components, we numerically solve the JEs for a "basis" of solutions (given by the solutions for \textit{mass-normalised} $\rho_{*i}$, $\phi_{*i}$, $\phi_\mathrm{h}$, $\phi_\mathrm{BH}$) and we project them. Then we exploit the linearity previously discussed, in addition with the linearity of the Satoh decomposition (eq. \ref{eq:satoh}) and of the projection operators, in order to recombine them in the post-processing tuning, when only the weight (mass, luminosity, anisotropy) coefficients are modified. This allows to build \textit{families} of models, by running the JASMINE code only once.

\section{Applications}

The applications of the analytical and numerical framework briefly illustrated in the previous section are numerous. It is now possible to reproduce, for example, different stellar kinematic components, such as counter rotating disks and kinematically decoupled cores. Furthermore, we can obtain luminosity-weighted total properties, by assuming a constant mass-to-light ratio different for each component. From a stellar population point of view, indeed, each component can be considered as a Simple Stellar Population, characterised by a certain age, metallicity, IMF, mass-to-light ratio (in agreement with Evolutionary Population Synthesis models, e.g. \cite[Maraston, 2005]{mar05}). The combination of different populations allows, for example, to study the systematic effects of IMF variations and mass-to-light ratio gradients on luminosity-weighted properties. The natural field of application of our study is the modeling of Early-Type Galaxies (ETGs), however the same approach can be also applied to modeling dwarf Spheroidal galaxies and Globular Clusters.

\begin{figure}[ht]
 \centering
 \includegraphics[scale=0.8]{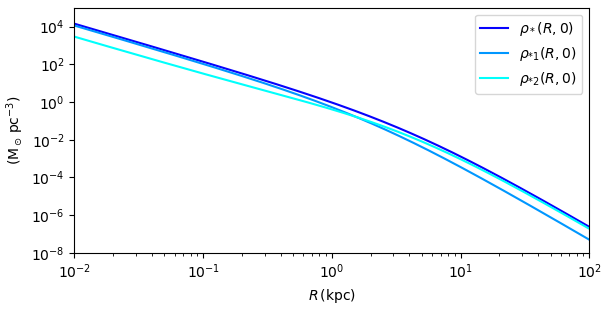}
 \caption{Density profiles on the equatorial plane of the two stellar components and total stars. The stellar component $\rho_{*2}$ is given by the difference of the two ellipsoidal Jaffe stellar models $\rho_*$ and $\rho_{*1}$. See Table \ref{tab:parameters} for the physical parameters adopted for this galaxy model.}
 \label{fig:rho}
\end{figure}

\begin{figure}[ht]
	\centering
	\includegraphics[scale=0.55]{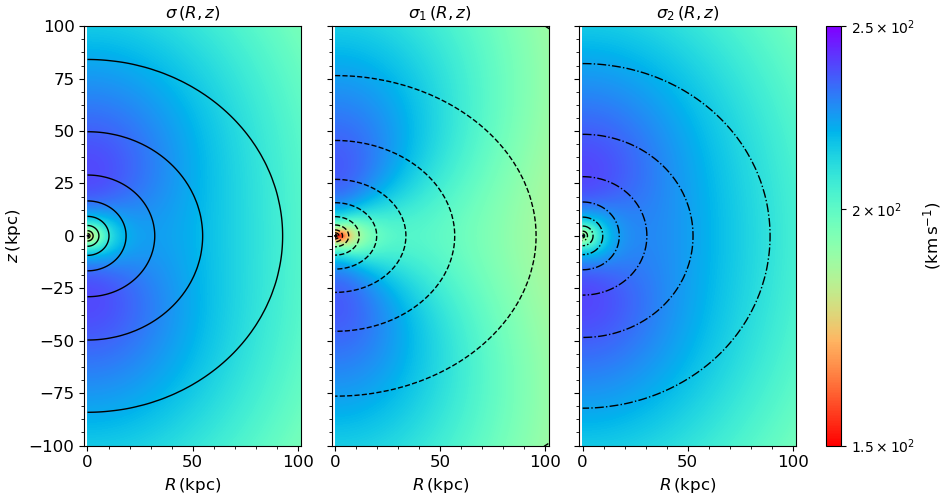}
	\caption{Radial (and vertical) components of the velocity dispersion of the total stars $\rho_*$ (left), and the two stellar components $\rho_{*1}$ (center) and $\rho_{*2}$ (right). In each panel, the contours represent the iso-densities of the corresponding component. See Table \ref{tab:parameters} for the physical parameters adopted for this galaxy model.}
	\label{fig:sigma}
\end{figure}

For illustrative purposes, we present a model of an elliptical galaxy, composed of two stellar populations, a DM halo and a central supermassive BH, whose parameters are reported in Table \ref{tab:parameters}. The total and first stellar components are modeled by ellipsoidal Jaffe density profiles, with a different flattening, mass and scale length. The second component is defined as the \textit{difference} of the two chosen distributions, $\rho_{*2}(R,z) = \rho_*(R,z) - \rho_{*1}(R,z)$, so that $M_{*2} = M_* - M_{*1}$, and in general it is \textit{not} ellipsoidal. In order to have a non-negative $\rho_{*2}(R,z)$ over the whole space, a \textit{non-negativity condition} is derived, constraining $q$, $q_1$, $M_{*1}/M_*$ and $r_{*1}/r_*$. In Figure \ref{fig:rho}, the stellar density profiles on the equatorial plane are shown. The DM halo is modeled by a spherical NFW density profile (\cite[Navarro, Frenk \& White, 1996]{NFW}), with given mass, scale radius and concentration. In Figure \ref{fig:sigma}, the radial (and vertical) velocity dispersions are shown for the two stellar components and for the total stars, as an example of the dynamics of a multi-component stellar system.
\begin{table}
	\centering
	\begin{tabular}{cccccccccc}
		\hline
		\multicolumn{2}{c}{Flattenings} &  \multicolumn{4}{c}{Masses} &
		\multicolumn{2}{c}{Scale lengths} & \multicolumn{2}{c}{Satoh parameters} \\
		$q$ & $q_1$ & $M_*$ & $M_{*1}/M_*$ & $M_\mathrm{h}/M_*$ & $M_\mathrm{BH}/M_*$ & $r_*$ & $r_{*1}/r_*$ & $k_1$ & $k_2$ \\
		\hline
		0.9 & 0.8 & $7.7 \cdot \mathrm{10^{10}\,M_\odot}$ & 0.3 & $10^2$ & $10^{-3}$ & 4.3 $\mathrm{kpc}$ & 0.5 & 1.0 & 0.0 \\
		\hline
	\end{tabular}
	\caption{Physical parameters of an illustrative model of an elliptical galaxy. This model is composed of two stellar populations, a DM halo and a central supermassive BH.}
	\label{tab:parameters}
\end{table}

\end{document}